\documentstyle[epsfig,eqsecnum,pre,aps,preprint]{revtex}

\begin{document} 
\tighten
\draft
\title{Spin block persistence at finite temperature}

\author{St\'ephane Cueille and Cl\'ement Sire}
\address{Laboratoire de Physique Quantique (UMR C5626 du CNRS),
Universit\'e Paul Sabatier\\
31062 Toulouse Cedex, France.\\}
\maketitle
\date{July 28,1997}

\begin{abstract}
We explore a new definition of the persistence exponent, measuring the
probability that a spin never flips after a quench of an Ising-like model at
a temperature $0<T<T_c$, while the usual definition only makes sense at
$T=0$. This probability is now defined for spin blocks, and a general scaling
for it, involving time and block linear size is introduced and illustrated by
extensive simulations.
\end{abstract}

Recent years have seen significant progress in the study of coarsening systems
\cite{brayrev94}. Our understanding of  phase ordering phenomena is now structured by
simple ideas such as dynamic scaling and universality. For instance, it is well
established that nonconserved order parameter dynamics are characterized by a
single length scale $L(t)\sim t^{1/z}$, and  that a nontrivial exponent
$\lambda$ appears in the scaling of the order parameter correlation function,
$\langle\varphi({\bf x},t')\varphi({\bf x},t)\rangle\sim [L(t)/L(t')]^\lambda$,
for $t'\gg t$. However, advances in this field have been to a large extent
boosted by progress in numerical simulations of lattice systems such as the
Ising model, while few analytical results are known except in $1d$. Therefore
surprises are still to be expected as one probes more and more complicated
correlations. 

Such a surprise came out  recently as much interest was devoted to the study of
the so-called  {\it persistence} probability
\cite{derrida94a,stauffer94,marcos95}. Consider the following simple question:
in a simulation of the Glauber dynamics of the Ising model at zero temperature,
what is the fraction of spins $p(t)$ which have never flipped  since the initial
time ? It turns out that $p$ exhibits a nontrivial algebraic decay $p(t)\sim
t^{-\theta}$. A quantity  such as $p(t)$ involves the whole history of the
system and is not easy to study analytically. Derrida {\it et al}
\cite{derrida95a} showed analytically that $\theta=3/8$ in $1d$, but in higher
dimensions, $\theta$ could only be determined by numerical simulations
\cite{derrida94a,stauffer94} or approximate methods
\cite{majumdar96a,majumdar96b,derrida96a}. More generally, the probability that
a stochastic physical quantity has never changed sign since the origin of its
evolution arises naturally in the context of nonequilibrium systems. Even for
simple scalar diffusion with zero mean random initial conditions, a nontrivial
algebraic decay  is found \cite{majumdar96b,derrida96a}.

For ferromagnetic systems, until very recently, persistence had only been
defined and studied at zero temperature  for a single spin (local order
parameter) \cite{derrida94a,stauffer94,derrida95a,majumdar96a,derrida96a} or
at $T_c$ for the total  magnetization (global order parameter), where it
yields a new independent critical exponent \cite{majumdar96c}. For $T>0$,
$p(t)$ decays exponentially due to thermal fluctuations, and the $T=0$
definition does not look very promising. However, there are good reasons to
be interested in a definition at a {\it finite temperature} $0<T<T_c$. First,
numerics, renormalization group arguments  or large-$N$ calculations
\cite{brayrev94} assess that finite (noncritical)  temperature  correlations
have the same scaling as zero temperature  correlations, with the same
$\lambda$. Thus it is natural to expect the same kind of universality to hold
for the persistence exponent, and it is worthwhile checking this point. In
addition, some discrepancies were found between the value of $\theta$ for
discrete (Ising) and continuous ($\varphi^4$) models. These discrepancies were
attributed to anisotropy effects in lattice systems at zero temperature, which
should be lowered at finite $T$.  Thirdly, certain important models  such as
the Ising model with conserved Kawasaki dynamics do not coarsen at zero
temperature, due to finite energy barriers, and must be simulated at $T>0$.
However  the corresponding continuous model (model B) exhibits coarsening at
zero temperature, and thus one needs a rule to extrapolate information from
finite temperature simulations using Kawasaki dynamics. 

In a very recent paper, Derrida \cite{derrida97a} proposed to study persistence
at finite temperature for nonconserved Ising and Potts models by comparing two
systems, $A$ and $B$, evolving with {\it the same thermal noise} from two
different  initial conditions: completely random for $A$, and all spins equal
to $+1$  for $B$ (fundamental state).  The idea is that $B$ experiences flips
solely due to thermal fluctuations in an ordered system. Thus a flip is
recorded only when a spin at the same site in both samples does not flip
simultaneously in $A$ and $B$. Derrida found in $d=2$ that the corresponding
persistence probability $p(t)$ decays algebraically with an exponent  close
to the value of $\theta$ at $T=0$. More extensive simulations performed by
Stauffer \cite{stauffer97} also suggest a temperature-independent  exponent
equal to $\theta$ in $d=2$, but significantly different in $d=3$  and $d=4$.
The value found in $d=3$ for $T>0$ is in good agreement with an 
approximate  continuous
theory  at $T=0$ \cite{majumdar96a} (see conclusion). Derrida's method is ingenious and
straightforwardly implemented, but it cannot be used to study conserved
dynamics, as system $B$ would not evolve with Kawasaki dynamics, and it is not
easy to generalize to a continuous field.

In this Letter, we propose a very natural method to study persistence at
finite temperature, namely ``block scaling'', which can be directly performed
on a single sample. The idea  stems from {\it \`a la  Kadanoff}
renormalization group ideas. At finite temperature, we consider the
persistence of coarse-grained spin variables obtained by  integrating the
order parameter (spin) on blocks. When the size $\l$ of the  blocks is
increased, the effective temperature flows to zero, which establishes a
connection with  the zero temperature dynamics. It is clear that this
definition also applies to continuous  models. We shall restrict ourselves
to nonconserved ferromagnetic  models (model A) with $L(t)\propto \sqrt{t}$,
but the same method can be used to study conserved models. 

Before considering finite temperature,  it is instructive to see how block
scaling works at $T=0$, for which two  persistence exponents can be defined:
$\theta$ for a single spin (local order parameter), and $\theta_0$ corresponding
to the probability  $p_0(t)$  that the  total magnetization (global order
parameter) has never changed sign \cite{majumdar96c,cornell97}. Majumdar {\it et
al} \cite{majumdar96c}  have shown the exact  result $\theta_0=1/4$ for the
Ising model in $1d$. Cornell and Sire  \cite{cornell97} performed direct
numerical simulations of $p_0(t)$ in $d=2$,  by recording the time when the
global magnetization first changes sign at $T=0$. This requires a very large
number of runs, which drastically limits the sample size ($L_{\text{max}}\sim
128$). In addition, finite size scaling  is not very conclusive, leading to a
large uncertainty on the value  of $\theta_0\approx 0.06\sim 0.11$. 

We now show that block scaling leads to a much easier determination of
$\theta_0$ at $T=0$, before moving to finite $T$. Let us consider  blocks of
size $\l$ and the probability $p_\l(t)$ that the  total magnetization of a block
has never changed sign since $t=0$ (we will use blocks with odd number $\l^d$ of
spins). For large time, when $L(t)\gg \l$, blocks behave as single spins and
$p_\l(t) \sim c_l t^{-\theta}$, $c_l$ being an increasing function of $\l$,
since obviously at large time $p_{\l'}(t)>p_\l(t)$ if $\l'>\l$. At early
times, when $L(t)\leq l^2$, the system effectively sees infinite blocks, and
$p_l(t)\propto t^{-\theta_0}$, where $\theta_0$ is the persistence exponent of
the {\it total magnetization} at $T=0$. Moreover, in the initial
configuration, the larger the blocks  the  smaller the  relative fluctuations
of the  magnetization around zero, therefore $p_{\l'}(t)<p_{\l}(t)$, for
$\l'>\l$.  The cross-over between the two regimes should occur at $t\propto
\l^2$.  These remarks lead us to the following large $l$ scaling,  
\begin{equation}\label{scaling}
p_\l(t)\sim \l^{-\alpha}f(t/\l^2)
\end{equation} 
where $f(x) \propto x^{-\theta_0}$ when $x\to 0$ and  $f(x) \sim x^{-\theta}$
when $x\to \infty$. For finite $t$,  $p_\l(t)$ must tend to a finite value for
$\l\to \infty$, equal to the  probability that the global magnetization never
changed sign. This requires $\alpha=2\theta_0$. Hence, computing $p_\l(t)$ for
several values of $\l$ makes it possible to determine $\theta_0$ by adjusting its
value to obtain the best data collapse.

\begin{figure}
\begin{center}
\epsfig{figure=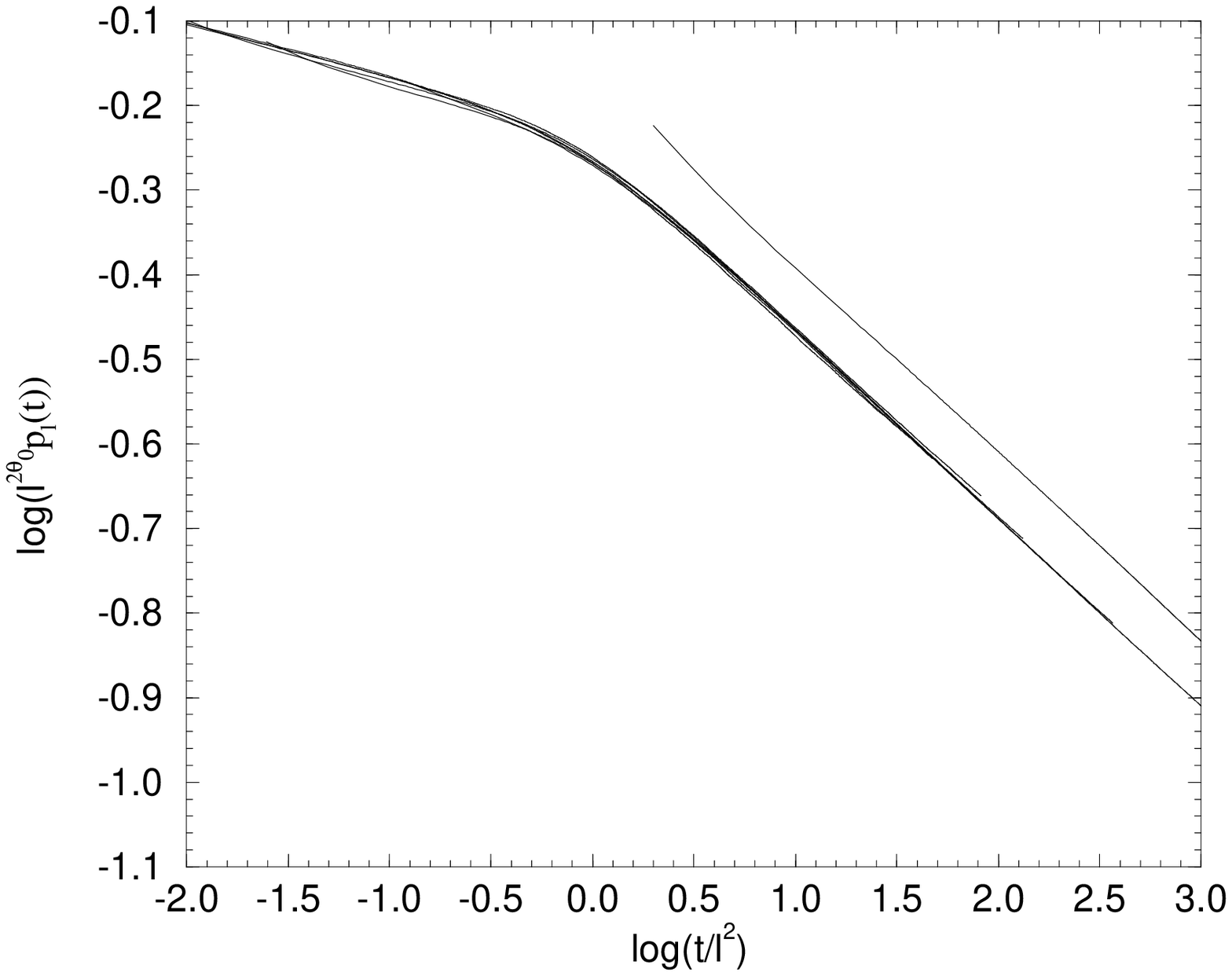, width=0.6\linewidth}
\caption{Block persistence at $T=0$ obtained from simulation of  the
nonconserved Ising model on a $2000^2$ lattice, for 
$\l=1,5,9,15,19,25,$ and $31$
(from bottom to top in the insert). $p_\l(t)$ decays as 
$t^{-\theta_0}$ at early
time and as  $t^{-\theta}$ at large time. Excellent scaling is then obtained
taking $\theta_0=0.09$.}
\label{zero}
\end{center}
\end{figure}

\begin{figure}
\begin{center}
\epsfig{figure=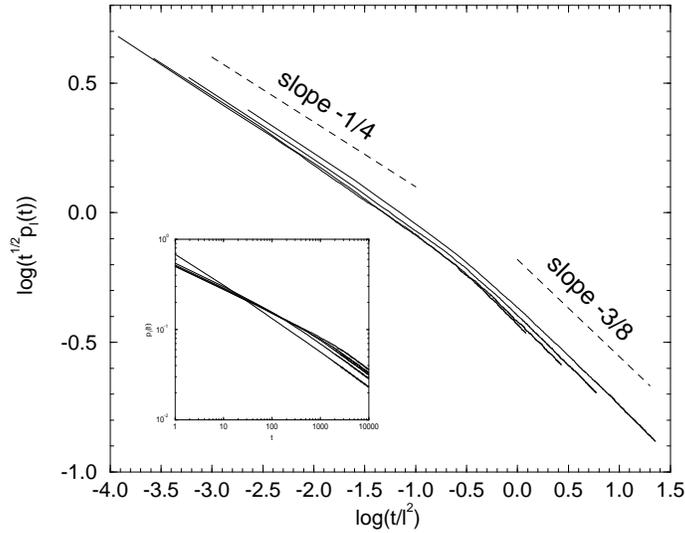, width=0.6\linewidth}
\caption{Similar scaling as in fig. 1, for a $1d$ spin chain (200000 spins, 10
samples), with block size $\l=1,21,41,61,91$ (from bottom to top in the
right part of the
insert). $\l=1$ is omitted in the scaling, and the data collapse improves as the
block size increases.} 
\label{1d}
\end{center}
\end{figure}

 To check this scaling, we simulated the $T=0$ Glauber dynamics
for the Ising model  in $d=2$ on a $2000^2$ lattice with blocks of linear size
1,5,9,15,19,25, and 31. 20 samples were averaged to obtain the final data
presented in fig. \ref{zero}. We find excellent scaling,  with $\theta_0=
0.09$.  Similar results were obtained in $1d$,  confirming the scaling
relation of Eq.  (\ref{scaling}) and the theoretical  value $\theta_0=1/4$
(fig. \ref{1d}).  Therefore, block scaling is a very convenient and reliable
method to determine $\theta_0$.

Now let us move to a finite temperature $0<T<T_c$ (not too close to $T_c$, a
case studied in a forthcoming paper \cite{cueille97e}).  The difficulty in
defining a persistence exponent  comes from the fact that a spin may flip due
to thermal fluctuations, leading  to an exponential decay $p(t)\sim \exp(-
t/{\tau})$.  Indeed, at $T=0$, a spin flips only when it is crossed by an
interface between a $+$ and a $-$ domain, whereas at finite temperature, the
dominating process at late time,  when the domains are large, is the flip of a
spin within a domain  due to thermal fluctuations. Therefore, at low
temperature,  it is natural from classical kinetics intuition to expect  an
Arrhenius law $\tau\sim \exp(-\Delta {\cal E}/T)$, where $\Delta {\cal E}$ is
the energy barrier to flip a spin (or a block) within an ordered domain. As
$T\to 0$, $\tau$ diverges and $p$ crosses over to a power law.

It is instructive to justify the Arrhenius law  from a random process viewpoint.
Let us consider a block of linear size $l$, and spin block variables
$\varphi_\l$. When  $L(t)$ is large enough,   the system can be considered
locally at equilibrium inside a domain, and, since there are no long-range
correlations, the relative fluctuation of  $\varphi_\l$ has the scaling $\Delta
\varphi_\l/\langle\varphi_l\rangle \propto \sqrt{T/\l^d}$.  Thus $p_\l(t)$ is
essentially the probability that a stationary  random process $X(t)$ with zero
mean and mean square fluctuation  $\langle X^2\rangle=T/\l^d$  crosses a barrier of
amplitude of order $1$. If $X(t)$ is Gaussian and Markovian, it is the solution
of a simple Langevin equation, with a Gaussian white noise $\eta(t)$ with
$\langle\eta(t)\eta(t')\rangle=2T/\l^d \delta(t-t')$. Then it is immediatly seen
that in exponential time $u=e^t$, $p_l(u)$ is the survival  probability of a
simple $1d$ random walker with diffusion coefficient $2T/\l^d$, starting from
$x=0$ with a moving  absorbing wall  at $x(u)\propto \sqrt{u}$. When the
amplitude of the fluctuation vanishes,  i.e. for small $T$ or large $\l$, this
survival probability can be evaluated by using the unperturbed solution of the
diffusion equation \cite{krapivsky}. At large $u$, $p_l(u)$ decays with a  power
law $p_l(u) \propto u^{-\beta}$ and $\beta \propto \sqrt{\l^d/T}
\exp(-C\l^d/T)$, where $C$ is a constant. Thus we recover the heuristic
Arrhenius law with $\tau\propto 1/\beta$.

The actual stochastic process $\varphi_l(t)$ is certainly non-Markovian.
However, for $\l$ much bigger than the equilibrium correlation length, it is
nearly Gaussian.  Moreover, its  correlator
$C(t)=\langle\varphi_l(t)\varphi_\l(0)\rangle$ can be bounded by two
Markovian exponential  correlators (because there is no long range correlation
in time  at equilibrium), and thus the Arrhenius law still holds with proper
constants inserted (although the power law in the prefactor may be modified)
\cite{krug97}. The important point is that the effective temperature
entering the Arrhenius law of the spin blocks is cut by a factor $\l^d$ and
that $\tau$ diverges very quickly when $\l$ is increased, leading to a fast
cross-over to the $T=0$ behavior. For $t\ll \tau$, $p_\l(t)$ is expected to
behave in the same way as for $T=0$. Finally, at finite temperature (not too
close to $T_c$, in the vicinity of which a different scaling arises
\cite{cueille97e}), we expect a scaling of the form $p_\l(t)\sim
\l^{-2\theta_0}f(t/\l^2)\exp[-t/\tau(\l,T)]$, involving two cross-over times
clearly visible in fig. 3, which shows the result of $2d$ simulations
performed at $T=2T_c/3$ on a $1000^2$ lattice, with $\l=1,3,5,7,9,11,13$.  The
exponential decay is clearly visible for $\l=1$ and $\l=3$. However, for larger
blocks, $\tau$ is bigger than the simulation time, and $p_\l(t)$ has the $T=0$
behavior, with a power law decay with exponent $\theta$  fully compatible with
the $T=0$ value ($\theta=0.22$), for $t\gg\l^2$, and a power law decay with
exponent $\theta_0$, for $t<\l^2$, just as expected. Figure \ref{t0.5} shows
the scaling with $\theta_0=0.09$  (for $\l=7,9,11,13$, and a slightly
smaller $T=T_c/2$ to eliminate the effect of the exponential cut-off). 


Thus, block scaling leads to a clear definition of $\theta$ at finite
temperature as the exponent of  the algebraic decay of the scaling function
$f(x)$. We find that in $2d$, the exponents $\theta$ and $\theta_0$ do not
depend on $T$ and are  equal to their $T=0$ value, in agreement with the
results obtained with Derrida's definition.  It is also very satisfactory to
observe that both scaling functions of fig. 1 and fig. 4 are identical up to a
multiplying factor, in a very similar way as what is known for the equal-time
two-point spin correlation function \cite{brayrev94}. 

We conclude with a look at the puzzling $3d$ case. Using the present block
method, we find an exponent $\theta_{T>0}$ consistent with the value
$\theta_{T>0}=0.26$ obtained by  Stauffer \cite{stauffer97} using Derrida's
definition, but different from the $T=0$ value $\theta_{T=0}=0.17$
\cite{stauffer94,majumdar96a}. In fact, it is well-known (although a precise
explanation  is still lacking) \cite{shore92,majumdar96a}, that the domain length
scale $L(t)$ does not grow as $t^{1/2}$ in $3d$, but as $t^{0.33}$, presumably
due to lattice effects. If we now express our general scaling as a function of
the more intrinsic $L(t)$ instead of time itself, we find that for $T=0$
and $T>0$  (in the latter case for $\l^2\ll t\ll \tau(l,T)$), both persistence
probabilities  decay as $p(t)\sim L(t)^{-\theta}$, with the same  $\theta
\approx 0.17/0.33\approx 0.26/0.5\approx 0.52$, in good agreement with the
theoretical prediction of \cite{majumdar96a}.

\vskip 0.3cm
\noindent We are very grateful to B. Derrida and S. Cornell for helpful
discussions.


\bibliographystyle{/h1/cueille/tex/bibtex/aps}
\bibliography{/h1/cueille/tex/these/coarsening}

\begin{figure}
\begin{center}
\epsfig{figure=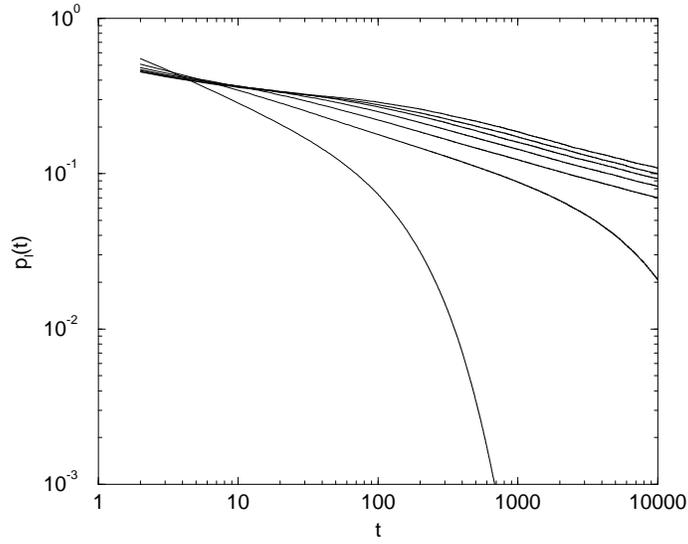, width=0.6\linewidth}
\caption{$p_l(t)$ for $T=2T_c/3$, and block sizes $\l=1,3,5,7,9,11,13$.}
\label{t2/3}
\end{center}
\end{figure}
\begin{figure}
\begin{center}
\epsfig{figure=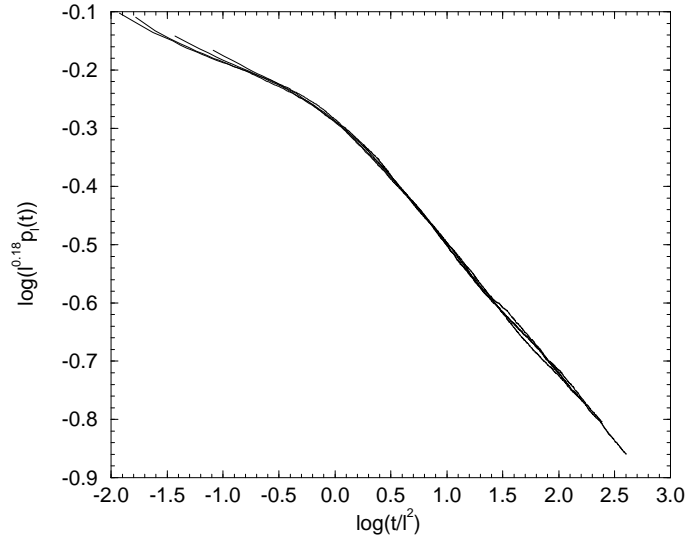, width=0.6\linewidth}
\caption{$p_l(t)$ expressed in scaling form for $T=T_c/2$, and block sizes
$\l=7,9,11,13$, using the same value for $2\theta_0=0.18$ as in the $T=0$ case.
Note the similarity with the $T=0$ scaling function of fig. 1.}
\label{t0.5}
\end{center}
\end{figure}
\end{document}